\documentclass[11pt]{article}

\usepackage{geometry}
\usepackage{setspace}
\usepackage{graphicx}
\usepackage{booktabs}
\usepackage{amsmath}
\usepackage{hyperref}
\usepackage{tcolorbox}
\usepackage{natbib}
\usepackage{authblk}

\title{
Recall, Risk, and Governance in Automated Proposal Screening for Research Funding: Evidence from a National Funding Programme
}

\author[1]{Chandan G. Nagarajappa\thanks{Corresponding author: \href{mailto:chandan.linus@gmail.com}{\texttt{chandan.linus@gmail.com}}}}
\author[1,2]{Moumita Koley}
\author[1]{Avinash Kumar}
\author[3]{Rabindra Panigrahy}
\author[3]{Pramod Kumar Arya}

\affil[1]{DST-Centre for Policy Research, Indian Institute of Science, Bengaluru, India - 560012}
\affil[2]{Research on Research Institute (RoRI), UK}
\affil[3]{Department of Science \& Technology, Ministry of Science \& Technology, India}

\date{}

\begin{document}
\maketitle

\begin{abstract}
Research funding agencies are increasingly exploring automated tools to support early-stage proposal screening. Recent advances in large language models (LLMs) have generated optimism regarding their use for text-based evaluation, yet their institutional suitability for high-stakes screening decisions remains underexplored. In particular, there is limited empirical evidence on how automated screening systems perform when evaluated against institutional error costs.

This study compares two automated approaches for proposal screening against the priorities of a national funding call: a transparent, rule-based method using term frequency–inverse document frequency (TF–IDF) with domain-specific keyword engineering, and a semantic classification approach based on a large language model. Using selection committee decisions as ground truth for 959 proposals, we evaluate performance with particular attention to error structure.

The results show that the TF–IDF-based approach outperforms the LLM-based system across standard metrics, achieving substantially higher recall (78.95\% vs.\ 45.82\%) and producing far fewer false negatives (68 vs.\ 175). The LLM-based system excludes more than half of the proposals ultimately selected by the committee. While false positives can be corrected through subsequent peer review, false negatives represent an irrecoverable exclusion from expert evaluation.

By foregrounding error asymmetry and institutional context, this study demonstrates that the suitability of automated screening systems depends not on model sophistication alone, but on how their error profiles, transparency, and auditability align with research evaluation practice. These findings suggest that evaluation design and error tolerance should guide the use of AI-assisted screening tools in research funding more broadly.
\end{abstract}

\textbf{Keywords:} research funding, proposal screening, TF-IDF, large language models, recall, false negatives, research governance

\section{Introduction}

Research funding agencies worldwide face a growing imbalance between the volume of submitted proposals and the administrative capacity available to evaluate them. While peer review remains the cornerstone of research funding decisions, the rapid increase in application volumes has placed substantial cognitive, temporal, and organisational burdens on programme officers responsible for initial screening. As a result, early-stage proposal screening has emerged as a critical, yet under-examined, component of contemporary research evaluation systems.

Internationally, several research funding organisations have begun experimenting with AI-assisted tools across different stages of the funding and evaluation process (\cite{Bommasani2021,(RoRI)2025}). These models rely on large-scale pretraining on textual corpora \citep{Radford2018}. Documented initiatives include the use of machine learning for reviewer matching at the Swiss National Science Foundation, AI-supported pre-review proposal scoring at the ‘La Caixa’ Foundation in Spain, automated tracking of funded research outputs at the Novo Nordisk Foundation in Denmark, and the use of AI-based methods to assess societal impacts of funded research by the Research Council of Norway. In the United Kingdom, large-scale experiments have explored whether AI systems could support post-publication research evaluation within the Research Excellence Framework. While these initiatives demonstrate growing global interest in algorithmic support for research funding, they have largely focused on discrete administrative functions rather than systematically examining the error profiles and institutional suitability of automated systems for early-stage proposal screening.

Early-stage screening differs fundamentally from peer review. Its purpose is not to assess scientific excellence in depth, but to ensure that proposals aligned with the objectives of a funding call progress to expert evaluation. In practice, screening decisions are made under conditions of limited time, incomplete information, and high uncertainty. Organisational research has long shown that such environments give rise to bounded rationality, where decision-makers rely on heuristics and satisficing strategies rather than exhaustive evaluation. In the context of research funding, these constraints increase the risk that potentially meritorious proposals are prematurely excluded from consideration.

In response to these pressures, funding agencies can explore computational approaches to assist in the selection of proposals. Advances in natural language processing (NLP), and particularly the emergence of large language models (LLMs), have generated optimism regarding their potential to support administrative decision-making involving unstructured text. However, despite growing interest, there is limited empirical evidence on whether such models are well-suited to the specific institutional requirements of early-stage screening in public research funding.

This study addresses this gap by empirically comparing two automated approaches to proposal screening against initial screening committee decisions: (i) a transparent, rule-based method using TF-IDF (a standard weighting scheme in information retrieval \citep{SaltonBuckley1988}) and domain-specific keyword engineering, and (ii) a semantic classification approach based on a large language model. Using a dataset of 959 proposals submitted to a national funding programme, we evaluate the performance of these approaches with particular attention to recall and false negatives.

We argue that evaluating screening systems solely on aggregate accuracy or precision obscures a critical feature of research funding contexts: the asymmetry of error costs. While false positives can be corrected through subsequent peer review, false negatives represent an irrecoverable institutional failure. By foregrounding this asymmetry, the paper contributes empirical evidence to ongoing debates in research evaluation regarding the appropriate design, deployment, and governance of AI-assisted screening systems.

These conditions closely reflect Herbert Simon’s concept of bounded rationality, which emphasises that organisational decision-makers operate under constraints of limited information, time, and cognitive capacity \citep{Simon1947,Simon1955}. In such environments, evaluation processes rely on satisficing strategies rather than exhaustive optimisation. Early-stage proposal screening can therefore be understood as an institutional response to bounded rationality, aimed at reducing cognitive load while preserving access to expert judgement.

In India, the integration of AI into research governance has begun to take shape primarily through public science administration institutions. The Department of Science and Technology (DST), through its policy and planning arm—the National Science, Technology and Innovation Management Information System (NSTMIS)—has initiated efforts to explore AI-assisted tools for research funding management, monitoring, and evaluation. These initiatives have focused on improving administrative efficiency, enhancing transparency, and supporting evidence-based decision-making across the research funding lifecycle. Within this institutional context, programme officers—typically scientists at the public funding agencies rather than a constituted expert committee—are required to process large volumes of proposals under severe time constraints, often relying on informal heuristics and proxy indicators in the absence of structured screening tools. This environment creates both the motivation and the governance challenge for developing automated screening systems that are aligned with public accountability, transparency, and institutional learning.

\section{Why Recall Matters in Early-Stage Proposal Screening}

Early-stage proposal screening functions as a safety mechanism rather than a final selection process. Funding agencies generally tolerate the inclusion of proposals that are marginal, weakly aligned, or ultimately unsuccessful, as such cases can be identified and rejected during subsequent peer-review stages. By contrast, proposals that are excluded during initial screening do not re-enter the evaluation pipeline, regardless of their underlying quality.

From an evaluation perspective, this asymmetry fundamentally alters how classification performance should be assessed. A false negative occurs when a proposal classified as out-of-scope by an automated system is subsequently judged as relevant or fundable by an expert committee. In research funding contexts, false negatives represent an irrecoverable loss of scientific opportunity and undermine procedural legitimacy, as applicants receive no substantive assessment of their work. False positives, by contrast, impose a comparatively low cost, typically in the form of additional reviewer workload.

From the perspective of bounded rationality (Simon, 1947; 1955), screening systems should be designed to minimise catastrophic errors under uncertainty, rather than optimise fine-grained discrimination at early decision stages \citep{Simon1955}.

Recall is therefore treated as the primary performance criterion for early-stage screening \citep{Powers2011}. Recall—the proportion of relevant proposals correctly retained—constitutes the primary performance criterion for early-stage screening systems. Optimising for precision or overall accuracy may produce superficially appealing results, but risks excluding a substantial share of meritorious proposals. In institutional terms, such conservatism is misaligned with the purpose of screening, which is to reduce administrative burden while preserving access to expert judgement.

This perspective aligns with established insights from research evaluation and public administration, which emphasise that decision-support systems operating under conditions of information overload should prioritise the avoidance of catastrophic errors over fine-grained discrimination. Automated screening tools should therefore be assessed not only in technical terms, but in relation to the organisational contexts and error tolerances within which they are deployed.

\section{Data and Methods}

\subsection{Dataset}

The analysis is based on 959 research proposals submitted to a national funding programme. Proposal objectives served as the primary textual input. Selection/screening committee decisions were used as ground truth, with 323 proposals selected and 636 rejected.

\subsection{TF-IDF and Keyword-Based Screening}

The TF-IDF approach employs manually curated, domain-specific keyword lists reflecting funding priorities related to policy-oriented over the technology-oriented research. Proposal texts were preprocessed using standard techniques, including lowercasing, punctuation removal, and stopword elimination. Proposals were deterministically classified based on keyword matches. TF-IDF vectors were computed for representation and exploratory analysis.

\subsection{LLM-Based Semantic Screening}

The LLM-based approach used a generative language model to assess whether proposal objectives aligned with predefined thematic categories. A structured prompt was used to elicit binary inclusion decisions. To enforce conservative filtering, deterministic exclusion rules were applied when proposals contained phrases explicitly outside the funding scope. 

The objective of this study is not to benchmark specific language models, but to evaluate the institutional suitability of semantic screening approaches relative to interpretable rule-based systems. While performance may vary across model sizes, architectures, or prompting strategies, the results highlight a structural risk associated with conservative semantic filtering in early-stage administrative screening contexts, namely the production of high false-negative rates.

\subsection{Evaluation Metrics}

Both methods were evaluated against selection committee decisions using accuracy, precision, recall, and F1 score. Given the institutional context, particular emphasis was placed on recall and the number of false negatives produced by each approach.

\section{Results}

This section presents the empirical comparison between the TF-IDF-based and LLM-based screening approaches. Results are organised to reflect the evaluation priorities of early-stage proposal screening, with particular emphasis on error structure and false negatives.

\subsection{Dataset Overview and Ground Truth Decisions}

Table~\ref{tab:dataset} summarises the composition of the dataset and the outcomes of the screening committee decisions used as ground truth. Of the 959 proposals analysed, 323 (33.7\%) were ultimately selected for further rounds of review which, means that they qualified for the next stage of evaluation, which involves peer review by domain experts, referred to as the expert committee.,  while 636 (66.3\%) were rejected. This distribution reflects the competitive nature of the funding programme and underscores the importance of effective early-stage screening mechanisms.

\begin{table}[h]
\centering
\caption{Dataset overview and expert committee decisions}
\label{tab:dataset}
\begin{tabular}{lcc}
\toprule
Category & Count & Percentage \\
\midrule
Total proposals analysed & 959 & 100.0\% \\
Selected by the committee & 323 & 33.7\% \\
Rejected by the committee & 636 & 66.3\% \\
\bottomrule
\end{tabular}
\end{table}

\subsection{Overall Classification Performance}

Table~\ref{tab:metrics} reports standard classification performance metrics for both screening approaches. The TF-IDF method outperformed the LLM-based approach across all metrics, achieving higher accuracy (80.92\% vs.\ 71.74\%), precision (68.92\% vs.\ 60.66\%), recall (78.95\% vs.\ 45.82\%), and F1 score (0.736 vs.\ 0.522).

While aggregate metrics provide a useful overview, they obscure important differences in error structure that are critical for institutional decision-making. For this reason, subsequent subsections focus explicitly on confusion matrices and error decomposition.

\begin{table}[h]
\centering
\caption{Overall classification performance of screening methods}
\label{tab:metrics}
\begin{tabular}{lcc}
\toprule
Metric & TF-IDF & LLM \\
\midrule
Accuracy & 80.92\% & 71.74\% \\
Precision & 68.92\% & 60.66\% \\
Recall & 78.95\% & 45.82\% \\
F1 Score & 0.736 & 0.522 \\
\bottomrule
\end{tabular}
\end{table}

\subsection{Confusion Matrix Analysis}

Tables~\ref{tab:cm_tfidf} and~\ref{tab:cm_llm} present the confusion matrices for the TF-IDF-based and LLM-based screening approaches, respectively. These tables make explicit the distribution of true positives, true negatives, false positives, and false negatives for each method.

The TF-IDF approach correctly identified 255 of the 323 committee-selected proposals, resulting in 68 false negatives. By contrast, the LLM-based approach identified only 148 of the committee-selected proposals, producing 175 false negatives. In other words, the LLM excluded more than half of the proposals that were ultimately judged as fundable by the expert committee.

\begin{table}[h]
\centering
\caption{Confusion matrix for TF-IDF-based screening}
\label{tab:cm_tfidf}
\begin{tabular}{lcc}
\toprule
 & \multicolumn{2}{c}{TF-IDF Classification} \\
Selection Committee Decision & Out & In \\
\midrule
Rejected (Out) & 521 (TN) & 115 (FP) \\
Selected (In) & 68 (FN) & 255 (TP) \\
\bottomrule
\end{tabular}
\end{table}

\begin{table}[h]
\centering
\caption{Confusion matrix for LLM-based screening}
\label{tab:cm_llm}
\begin{tabular}{lcc}
\toprule
 & \multicolumn{2}{c}{LLM Classification} \\
Selection Committee Decision & Out & In \\
\midrule
Rejected (Out) & 540 (TN) & 96 (FP) \\
Selected (In) & 175 (FN) & 148 (TP) \\
\bottomrule
\end{tabular}
\end{table}

\subsection{Error Decomposition and Institutional Risk}

To further clarify the implications of these results, Table~\ref{tab:errors} decomposes classification errors by type. While the LLM-based approach produced slightly fewer false positives (96 vs.\ 115), it generated more than twice as many false negatives as the TF-IDF approach (175 vs.\ 68). The total error rate of the LLM-based system (28.26\%) was also substantially higher than that of the TF-IDF system (19.08\%).

From an institutional perspective, this error profile is critical. False negatives correspond to proposals that are irreversibly excluded from expert evaluation, whereas false positives can be corrected at later stages. The error decomposition therefore reinforces the conclusion that the TF-IDF approach is better aligned with the functional role of early-stage screening.

\begin{table}[h]
\centering
\caption{Error decomposition by screening method}
\label{tab:errors}
\begin{tabular}{lcc}
\toprule
Error type & TF-IDF & LLM \\
\midrule
False Positives & 115 & 96 \\
False Negatives & 68 & 175 \\
Total errors & 183 & 271 \\
Error rate & 19.08\% & 28.26\% \\
\bottomrule
\end{tabular}
\end{table}

\subsection{Inter-Method Agreement}

Finally, Table~\ref{tab:agreement} reports the level of agreement between the two automated screening approaches. The methods agreed on 645 proposals (67.26\%), including 150 cases where both selected a proposal and 495 cases where both rejected it. However, disagreement occurred in nearly one-third of cases (32.74\%).

Notably, in 220 cases (22.94\%), the TF-IDF method selected a proposal that the LLM rejected, compared to only 94 cases (9.80\%) where the LLM selected a proposal rejected by TF-IDF. This asymmetry further illustrates the conservative nature of the LLM-based approach and its tendency to exclude proposals that the TF-IDF system retains.

\begin{table}[h]
\centering
\caption{Agreement between TF-IDF and LLM screening outcomes}
\label{tab:agreement}
\begin{tabular}{lcc}
\toprule
Agreement category & Count & Percentage \\
\midrule
Both methods agree & 645 & 67.26\% \\
Both select (In) & 150 & 15.64\% \\
Both reject (Out) & 495 & 51.62\% \\
Methods disagree & 314 & 32.74\% \\
TF-IDF In, LLM Out & 220 & 22.94\% \\
TF-IDF Out, LLM In & 94 & 9.80\% \\
\bottomrule
\end{tabular}
\end{table}

\section{Discussion}

This study set out to evaluate the suitability of different automated approaches for early-stage research proposal screening, with particular attention to institutional error costs and governance considerations. Rather than treating proposal screening as a purely technical classification task, the analysis foregrounds the administrative context in which such systems operate. The results demonstrate that differences in model performance are not merely quantitative, but have qualitatively distinct implications for research evaluation practice.

\subsection{Error Asymmetry and Administrative Risk}

The overall performance metrics reported in Table~\ref{tab:metrics} indicate that the TF-IDF-based approach outperforms the LLM-based system across accuracy, precision, recall, and F1 score. However, aggregate metrics alone are insufficient for evaluating screening systems in funding contexts. As shown by the confusion matrices in Tables~\ref{tab:cm_tfidf} and~\ref{tab:cm_llm}, the most consequential differences between the two approaches lie in their error structure.

The TF-IDF method produces 68 false negatives, whereas the LLM-based approach produces 175 (Table~\ref{tab:errors}). In practical terms, this means that the LLM excludes more than half of the proposals ultimately selected by the selection committee. From an institutional perspective, this represents a substantial administrative risk. False negatives correspond to proposals that are irreversibly excluded from expert evaluation, while false positives can be corrected during subsequent peer-review stages. As a result, screening systems that prioritise conservative exclusion risk undermine the core purpose of early-stage evaluation.

As highlighted earlier, this asymmetry can be interpreted through the lens of Herbert Simon’s concept of bounded rationality. Under conditions of limited time, information, and cognitive capacity, decision-making is better oriented toward minimising the risk of severe errors than toward fine-grained optimisation. In the context of research funding, the most consequential error is not the inclusion of a few weak proposals, which can be removed in expert review, but the premature exclusion of proposals that are genuinely meritorious and warrant further consideration.  The TF-IDF approach, by retaining a larger share of committee-selected proposals, aligns more closely with this institutional requirement.

\subsection{Conservatism and Disagreement Between Screening Approaches}

Further insight into the behaviour of the two screening systems is provided by the inter-method agreement analysis in Table~\ref{tab:agreement}. Although the methods agree on approximately two-thirds of proposals, disagreement occurs in nearly one-third of cases. Importantly, these disagreements are not symmetric. In 220 cases, the TF-IDF method selects proposals that the LLM rejects, compared to only 94 cases in which the LLM selects proposals rejected by TF-IDF.

This pattern indicates that the LLM-based approach adopts a more conservative screening posture, systematically excluding proposals that the TF-IDF system retains. While such conservatism may be desirable in contexts where false positives are costly, it is misaligned with the functional role of early-stage screening in research funding. As Table~\ref{tab:errors} demonstrates, this conservatism translates directly into a substantially higher false negative rate, increasing the risk of irrecoverable exclusion.

\subsection{Interpretability, Accountability, and Institutional Fit}

Beyond performance metrics, the results raise broader questions about the governance of automated screening systems. Public funding agencies operate under expectations of transparency, accountability, and procedural legitimacy. Evaluation processes must be defensible not only internally, but also to applicants and external stakeholders.

The TF-IDF-based approach offers a high degree of interpretability, as decisions can be traced directly to explicit keywords and rules. This traceability facilitates auditing, iterative refinement, and organisational learning. By contrast, the LLM-based approach relies on opaque internal representations and prompt-dependent behaviour, making systematic auditing more difficult even when justificatory outputs are provided. 

Although domain-specific keyword engineering requires initial expert input, this cost is front-loaded and supports iterative refinement through auditability, in contrast to opaque model behaviour that is harder to diagnose and correct once deployed.

As emphasised in institutional analyses of collective decision-making, transparency and feedback mechanisms are central to organisational learning and legitimacy \citep{Ostrom1990,Ostrom2005}. Elinor Ostrom’s work on institutional design highlights the importance of transparency and feedback mechanisms in sustaining trust and learning in collective decision-making systems. In the absence of traceable decision pathways, organisations struggle to reflect on outcomes, refine criteria, or communicate clear expectations. From this perspective, the interpretability of the TF-IDF approach constitutes a significant institutional advantage that is not captured by performance metrics alone.

Taken together, the results reported in Tables~\ref{tab:metrics} through~\ref{tab:agreement} suggest that model sophistication alone is a poor guide for selecting screening tools in research evaluation contexts. Instead, the suitability of automated systems depends on how their error profiles, transparency, and operational characteristics align with the institutional realities of public research funding.

\section{Conclusion}

This study provides empirical evidence that a transparent, TF-IDF-based screening system with domain-specific keyword engineering can outperform a general-purpose large language model in early-stage research proposal screening. In particular, the TF-IDF approach substantially reduces false negatives, ensuring that a greater proportion of meritorious proposals reach expert review.

By explicitly foregrounding error asymmetry, the findings highlight the importance of aligning evaluation metrics and system design with the institutional realities of research funding. In screening contexts where exclusion is irreversible, recall constitutes a more appropriate performance criterion than precision or overall accuracy. More complex models do not necessarily yield better outcomes when their error profiles and governance characteristics are misaligned with organisational needs.

The results suggest that responsible innovation in research evaluation does not require increasingly sophisticated AI systems, but rather careful attention to transparency, auditability, and institutional fit. For funding agencies facing growing proposal volumes, well-engineered, interpretable NLP tools may offer a more reliable and defensible foundation for automated screening than general-purpose language models.

\section*{Acknowledgements}
The authors acknowledges the use of generative AI tools to assist with language editing and the correction of grammatical issues in earlier drafts of this manuscript. All substantive ideas, analyses, interpretations, and conclusions are the authors' own, and the authors takes full responsibility for the content of the paper.

\bibliography{references}

\end{document}